\documentclass[floatfix,twocolumn,aps,prl,showpacs,amsmath,amssymb]{revtex4-1}
\usepackage[latin1]{inputenc}
\usepackage[dvips]{graphicx}

\usepackage{dcolumn}
\usepackage{bm}
\usepackage{dsfont}
\usepackage{color}
\usepackage{url}
\usepackage[colorlinks=true ,citecolor=blue]{hyperref}
\usepackage{amsmath,amssymb,esint}
\usepackage[table]{xcolor}

\definecolor{Nathanblue}{rgb}{0.,0.24,0.51}

\newcommand{\be}{\begin{equation}}
\newcommand{\ee}{\end{equation}}
\newcommand{\bq}{\begin{eqnarray}}
\newcommand{\eq}{\end{eqnarray}}

\begin{document}
	
	\title{Noncommutative Geometry and Deformation Quantization\\ in the Quantum Hall Fluids with Inhomogeneous Magnetic Fields}

	\author{Giandomenico Palumbo}
	\affiliation{School of Theoretical Physics, Dublin Institute for Advanced Studies, 10 Burlington Road,
		Dublin 4, Ireland}
	
	\date{\today}
	
	\begin{abstract}
		
		\noindent It is well known that noncommutative geometry naturally emerges in the quantum Hall states due to the presence of strong and constant magnetic fields. Here, we discuss the underlying noncommutative geometry of quantum Hall fluids in which the magnetic fields are spatially inhomogenoeus. We analyze these cases by employing symplectic geometry and Fedosov's deformation quantization, which rely on symplectic connections and Fedosov star-product. Through this formalism, we unveil some new features concerning the static limit of the Haldane's unimodular metric and the Girvin-MacDonald-Platzman algebra of the density operators, which plays a central role in the fractional quantum Hall effect.
		
	\end{abstract}
	
	\maketitle
	\section{Introduction}
\noindent Noncommutative geometry represents not only an important theoretical approach in high energy physics \cite{Connes,Seiberg,Szabo2,Lizzi, Landi, Balachandran,Vassilevich,Gubitosi}, but it has
	found in the quantum Hall effect an experimentally testable framework, in which some of its mathematical predictions have important physical implications.
	This is the case for the magneto-roton excitation, a spin-2 collective mode (massive graviton), which is related to the Girvin-MacDonald-Platzman (GMP) algebra \cite{Girvin,Papic,Golkar,Papic2}. 
	This is an infinite-dimensional sine-algebra \cite{Zachos} of the projected density operators in the lowest Landau Level (LLL) and is dual to the $W_{\infty}$ algebra of quantum area-preserving diffeomorphisms in two space dimensions \cite{Karabali,Cappelli,Cappelli2}. 
	The incompressibility of the fractional quantum Hall fluids is deeply related to the emergence of a noncommutative geometry, in which the fundamental Planck length is replaced by the magnetic length.
	For this reason, noncommutative geometry and noncommutative field theories have been largely employed to study several aspects of the quantum Hall states \cite{Bellissard,Susskind,Polychronakos,Hasebe,Karabali2,Prodan,Szabo3,Cobanera,Cappelli3,Senthil,Son}.
	Complementary to the operational framework, noncommutative geometry can be also formalized through the deformation quantization approach, which is based on the Moyal star-product and symplectic geometry \cite{Szabo2,Sternheimer,Vitale}. 
	In fact, the incompressibility of the quantum Hall fluids is related to quantum symplectomorphisms and
	 the GMP algebra and the edge states of the quantum Hall fluids can be directly derived by applying the Moyal brackets (which are built from the Moyal star-product) to density operators \cite{Karabali,Cappelli3,Szabo3,Stone}.
	Although deformation quantizaton in the quantum Hall effect is well understood in the case of strong and constant magnetic fields, the situation is more involved for inhomogeneous magnetic fields, which in general give rise to dispersion of the Landau levels and to wave functions that are not holomorphic.
	It is possible to show that LLLs and holomorphicity are well posed even in curved space with
	inhomogeneous magnetic fields, if the magnetic field is proportional to the Kähler form \cite{Klevtsov,Wiegmann,Ledwith}. 
	Although this complex-geometric approach is powerful and in part related to the geometric quantization of the quantum Hall states \cite{Duval,Nair2,Klevtsov,Ozawa4,Oblak,Oblak2}, in this paper we will main employ the deformation quantization approach that does not require the identification of any complex structure.

	\noindent The main goal of this work is to unveil some new features related to the GMP algebra and the static limit of the Haldane's unimodular metric \cite{Haldane,Sondhi} for static and weak inhomogeneous magnetic fields. We will employ symplectic geometry and Fedosov's deformation quantization, which require the introduction of symplectic connections together with a covariant generalization of the Moyal star-product known as Fedosov star-product \cite{Fedosov,Gelfand} (see also Refs \cite{Moore,Ahn}, for a recent application of symplectic connections to condensed matter systems and Refs \cite{Karabali4,Karabali5,Nair4} where a start-product, similar to that one by Fedosov, is introduced in the context of the quantum Hall effect). Although the Fedosov's deformation quantization has been already employed in the high-energy-physics context to study (curved) noncommutative geometries with space-dependent deforming parameters \cite{Dobrski,Blaschke}, in our case, this parameter will be simply interpreted as a spacially-varying magnetic length.
	\section{Noncommutative geometry with constant magnetic fields}
	\noindent	We start reviewing the well known case of electrons in the plane under the influence of a strong constant magnetic field $B$. Here, the electrons rapidly rotate around the centers of cyclotronic orbits known as guiding centers $\tilde{X}_i$. There are two different ways to see the emergent noncommutative geometry: by starting from a semiclassical action for the non-relativistic electrons and then employing the quantization in the ultrastrong magnetic field regime or by directly projecting onto the LLL and taking the ultrastrong-magnetic field limit \cite{Cobanera}. Here, we adopt the first approach, which can be more easily generalized to the case of (weakly) inhomogeneous magnetic fields as we will see in the next section.
	The Lagrangian for a non-relativistic electron on the plane in presence of a magnetic field is given by
	\begin{equation}
		L=\frac{1}{2}\left( \dot{X}^2 + \dot{Y}^2 \right)+e B Y \dot{X} - V(X,Y),
	\end{equation}
where $V(X,Y)$ is an external potential, $X,Y$ the spatial coordinates, $m$ and $e$ are the mass and charge of the electron, respectively and we have chosen the Landau gauge for the gauge connection $A_{\mu}=(0,-B Y, 0)$, with B a constant magnetic field and $\mu=\{t,X,Y\}$. Here, for simplicity, we have fixed the effective velocity of light to one, $c=1$. In the limit of large B field, we can completely
neglect the first term, namely the kinematic term such that the effective Lagrangian $\bar{L}$ for an electron in the ultra-strong magnetic regime only depends on the second and third term. In this way, we have now a Lagrangian that is linear in velocity and the corresponding quantization can be derived through the Faddeev-Jackiw approach \cite{Jackiw}. The Hamiltonian is given by
\begin{equation}
	H= p_{x} \dot{X}-\bar{L}=V(X,Y),
\end{equation}
where $p_{x}=\partial \bar{L}/ \partial \dot{X}=e B Y$ is the only non-zero momentum variable. This is the natural regime in which the fractional quantum Hall effect emerges, being the effective Hamiltonian of the fractional quantum Hall states entirely determined by the many-body interactions encoded in V.
Upon quantization, the operators $\hat{X}$ and $\hat{p}_x$ satisfy the Heisenberg commutation relation $[\hat{X}, \hat{p}_x]=i \hbar$ such that
\begin{equation} \label{noncommutative}
	[\hat{X}, \hat{Y}]= i l^2_B,
\end{equation}
	where $l_B=\sqrt{\hbar/e B}$. This relation represents the noncommutative geometry of the plane.  A similar expression holds for the guiding center operators defined by
		\begin{equation}
		\tilde{X}=\hat{X}-\frac{l_B^2}{\hbar}\hat{\pi}_y, \hspace{0.3cm} \tilde{Y}=\hat{Y}+\frac{l_B^2}{\hbar}\hat{\pi}_x,
	\end{equation}
where $\hat{\pi}_i = \hat{p}_i + e A_i$ are the covariant momentum operators 
		that satisfy the commutation relation
		\begin{equation}
				[\hat{\pi}_x, \hat{\pi}_y]= -i \frac{\hbar^2}{l_B^2}.
		\end{equation}
	Notice that although here we have considered the symmetric gauge instead of the Laundau gauge, it is well known the physics does not depend on the gauge fixing. Intuitively, we can understand the similar behaviour between the standard coordinates of the electron and the corresponding guiding centers in the ultra-strong magnetic limit because the stronger the B field the smaller is the radius of the Landau orbits. From $[\tilde{X}, \tilde{Y}]= i l^2_B$, we can now derive the GMP algebra for the projected density operators $\bar{\rho}(q_x, q_y)$
		\begin{equation}
		\bar{\rho}(p_x,p_y) = e^{i (p_x \tilde{X}+p_y \tilde{Y})},
	\end{equation}
	By employing the Baker-Campbell-Hausdorff formula to write \cite{Szabo3}
		\begin{align}
	e^{i (p_x \tilde{X}+p_y \tilde{Y})}e^{i (q_x \tilde{X}+q_y \tilde{Y})}=\nonumber \\
	 e^{-\frac{i}{2}l_B^2(p_x q_y - p_y q_x)} e^{i ((p_x+q_x) \tilde{X}+(p_y+q_y) \tilde{Y})}
	\end{align}
	we finally obtain
		\begin{eqnarray}
		[\bar{\rho}(p_x,p_y), \bar{\rho}(q_x,q_y)]=  \hspace{0.8cm}\nonumber   \\
		2 i \sin \left[(p_x q_y - p_y q_x)l_B^2/2\right] \bar{\rho}(p_x+q_x, p_y+q_y),
	\end{eqnarray}
which is the GMP algebra that plays a central role in the stability of the fractional quantum Hall states and in their incompressibility. In fact, this algebra is related to the quantum area-preserving diffeomorphisms on the plane while $\tilde{X}_i$ are nothing but the generators of the magnetic translations. Among the other things, through the GMP algebra is possible to derive a collective mode in the bulk of fractional quantum Hall states known as magneto-roton, which is a spin-2 massive propagating mode \cite{Girvin,Papic,Golkar}.
It is also possible to start from this algebra and derive the anticommutation relation of the guiding centers \cite{Sondhi2}. We can in fact take first the the long-wavelength limit of the above algebra by getting
\begin{eqnarray}
	[\bar{\rho}(p_x,p_y), \bar{\rho}(q_x,q_y)]\approx  \hspace{0.8cm}\nonumber   \\
	i  (p_x q_y - p_y q_x)l_B^2 \bar{\rho}(p_x+q_x, p_y+q_y),
\end{eqnarray}
and then define $\tilde{X}_i$ as follows
\begin{eqnarray}
	\tilde{X}_i = -i \lim_{p_i\rightarrow0} \partial_p \bar{\rho},
\end{eqnarray}
such that $[\tilde{X}, \tilde{Y}]= i l^2_B$.
Following Refs \cite{Karabali,Cappelli3,Szabo3,Stone}, we show now that the GMP algebra can be derived by implementing the Moyal brackets to the density operators. Moyal star-product (and consequently Moyal brackets) is associative but noncommutative and is at the core of deformation quantization, which is a powerful approach in noncommutative geometry \cite{Szabo2,Sternheimer,Vitale}. 
In fact, from the noncommutative algebra induced by the operators $\hat{X}_i$ is possible to pass to the commutative and associative algebra of smooth functions (e.g. $f$ and $g$) that depend on the real variables $X_i$, which are equipped with a Moyal star-product $\star$, namely
\begin{equation}\label{MoyalB}
\{f(X,Y), g(X,Y)\}_M= f \star g - g \star f,
\end{equation}
with
\begin{equation}
f \star g = f(X,Y) \exp\left(\frac{i l_B^2}{2}  \epsilon^{ij} \overleftarrow{\partial}_i \overrightarrow{\partial}_j\right) g(X,Y).
\end{equation}
To leading order in $l_B$, i.e. the deforming parameter, the Moyal brackets in Eq. (\ref{MoyalB}) become the classical Poisson brackets.
For 
\begin{equation}
f(X,Y)=X, \hspace{0.3cm} g(X,Y)=Y, 
\end{equation}
we indeed obtain
\begin{equation}
\{X, Y\}_M= i l^2_B,
\end{equation}
which shows that for the noncommutative plane, the Moyal star-product approach is equivalent to the
operatorial approach based on the commutation relations in Eq.(\ref{noncommutative}).
For real-space density operators
\begin{equation}
\hat{\rho}(X,Y)=\hat{\psi}^{\dagger}\hat{\psi}, 
\end{equation}
with $\hat{\psi}$ the fermionic wave-function operator, we can consider the corresponding Fourier-transformed density operator
\begin{equation}
\hat{\rho}(p_x,p_y)= \int dX dY\, \hat{\rho}(X,Y)\, e^{i (p_x X+p_y Y)},
\end{equation}
such that the commutation relations associated to algebra of the momentum-space density operators are simply given by the Moyal brackets of two plane waves, i.e.
\begin{equation}\label{planewaves}
\rho(p_x,p_y)\equiv e^{i (p_x X+p_y Y)}, \hspace{0.3cm}\rho(q_x,q_y)\equiv e^{i (q_x X+q_y Y)},
\end{equation}
and by defining
\begin{align}
\langle \{\rho(p_x,p_y), \rho(q_x,q_y)\}_M\rangle\equiv \nonumber \\
 \int dX dY\, \hat{\rho}(X,Y)\, \{\rho(p_x,p_y), \rho(q_x,q_y)\}_M,
\end{align}
we finally obtain
\begin{align}
[\hat{\rho}(p_x,p_y), \hat{\rho}(q_x,q_y)]=\langle\{\rho(p_x,p_y), \rho(q_x,q_y)\}_M \rangle= \nonumber \\
 2 i \sin \left[(p_x q_y - p_y q_x)l_B^2/2\right] \hat{\rho}(p_x+q_x, p_y+q_y),
\end{align}
which coincides with the GMP algebra. This is a very general result for fractional quantum Hall states for which the filling fraction is implicitly encoded in the density.
Before concluding this section, we want to remark that the above relation is related to a more general result that involves classical averages of generic functions $f(X,Y)$ and $g(X,Y)$ on the plane. In fact, by defining the averages of $f$ and $g$ as 
\begin{align}
\langle f \rangle= \int dX dY\, \hat{\rho}(X,Y)\, f(X,Y), \nonumber \\ \langle g \rangle= \int dX dY\, \hat{\rho}(X,Y)\, g(X,Y),
\end{align}
it is then possible to show that \cite{Karabali,Szabo3,Stone}
\begin{equation}
[\langle f \rangle, \langle g \rangle] =\langle \{f, g\}_M \rangle,
\end{equation}
which represents one of the most explicit and direct manifestations of Moyal brackets and deformation quantization in lower-dimensional physical systems. 
\section{Inhomogeneous magnetic fields and Fedosov's deformation quantization}
\noindent Here, we consider the case of a weakly inhomogeneous magnetic field $B(X)$, in which the electrons keep on rapidly rotating around the guiding centers $\tilde{X}_i$ that slowly drift in the two-dimensional space. This implies that the cyclotron frequency becomes a space-dependent function, i.e. $\omega_{B}(X) \simeq e B(X)/mc$. We can then distinguish two kinds of motion of the system: the slow drift of the guiding center and the fast rotation around the guiding center. Bear in mind that in weakly inhomogeneous magnetic fields, 
the magnetic length is considered small with respect to the length scale over which the magnetic fields vary on the 2D space.
In the generalized Landau gauge $A_{\mu}=(0,-B(X) Y, 0)$, we have 
\begin{equation}
\hat{p}_x=e \hat{A}_x \equiv e \hat{B}(\hat{X}) \hat{Y},
\end{equation}
with $[\hat{X}, \hat{p}_x]= i \hbar$ such that
\begin{align}
 [\hat{X}, \hat{Y}]= \frac{i\hbar}{e} (\hat{B}(\hat{X}))^{-1},
\end{align}
by assuming that $\hat{X}$ and $B(\hat{X})$ commute, with the latter an invertible operator. Moreover, because the two-dimensional inversion symmetry given by
\begin{align}
X\rightarrow -X, \hspace{0.3cm} Y\rightarrow -Y,
\end{align}
has been recently recognized by Haldane \cite{Haldane2} as the fundamental discrete symmetry of the fractional quantum Hall states, $B(\hat{X})$ needs to be even under inversion symmetry (in other words, inversion symmetry selects the weakly inhomogeneous magnetic fields that are compatible with the fractional quantum Hall effect).
Here, the corresponding noncommutative operator 
\begin{equation}
\hat{\theta}^{ij}(\hat{X})=\epsilon^{ij}\,\frac{\hbar}{e}\, (\hat{B}(\hat{X}))^{-1}
\end{equation}
with $\{i,j\}=\{X,Y\}$, is space-dependent and identified as a space-varying magnetic length. In noncommutative geometry, space-dependent noncommutative parameters have been analyzed in Refs \cite{Dobrski,Blaschke}, while 
the above expression for the quantum Hall effect has been already derived by Maraner \cite{Maraner} through the operatorial approach by introducing generalized guiding center operators with a symmetric gauge. This implies that noncommutative geometry of generalized guiding centers should have some non-trivial effects on the GMP algebra.\\ 
In order to show the physical consequences of the above expression for the density operators, we here adopt the Fedosov's deformation quantization approach \cite{Fedosov,Gelfand}, in which Fedosov star-product and symplectic connections play a central role. 
 We start by noticing that the Weyl symbol of $\hat{\theta}^{ij}(\hat{X})$ at leading order can be identified with the real function $\theta^{ij}(X)$, which represents a Poisson tensor because it trivially satisfies the Poisson-Jacobi identity \cite{Blaschke}
 \begin{equation}
 \theta^{ij}\partial_j \theta^{kl}+\theta^{kj}\partial_j \theta^{li}+\theta^{lj}\partial_j \theta^{ik}=0,
 \end{equation}
with $\{i,j,k,l\}=\{X,Y,Z\}$. Because we are also assuming that $\theta^{ij}$ is invertible for any $X$, we can define its inverse
\begin{equation}\label{omega2}
\omega_{ij}= \epsilon_{ij}\,\frac{e}{\hbar} B(X),
\end{equation}
which can be seen as a symplectic two-form that identifies a symplectic manifold. Thus, at classical level, the area-preserving diffeomorphisms induced by magnetic fields on the plane can be mapped to two-dimensional symplectomorphisms. Symplectic geometry plays a central role in the Fedosov's deformation quantization.
Similarly to Riemmanian geometry that deals with Christoffel symbols (i.e. metric connections) built from a symmetric metric tensor, in the symplectic case, it is possible to define (torsionless) symplectic connections $\Gamma^{i}_{jk}$ (also known as Fedosov connections) from a generic non-constant simplectic two-form as follows
\begin{equation}
\Gamma^{k}_{ij}= \frac{1}{3} \left(\partial_i \omega_{kj}+\partial_j \omega_{ki}\right).
\end{equation}
Symplectic connections are symmetric in the lower indices, i.e. $\Gamma^{i}_{jk}=\Gamma^{i}_{kj}$ and can be employed to build covariant derivatives
\begin{equation}
D_i v^j = \partial_i v^j + \Gamma^j_{k i} v^k,
\end{equation}
where $v^j$ is a tangent vector field and symplectic Riemann tensors 
\begin{equation}\label{Riemann}
R^i_{jkl}=\partial_k \Gamma^i_{jl}-\partial_l \Gamma^i_{jk}+ \Gamma^i_{mk}\Gamma^m_{jl}
-\Gamma^i_{ml}\Gamma^m_{jk}
\end{equation}
on symplectic manifolds. Similar to the Riemannian case, a single symplectic covariant derivative acts trivially on a real function, i.e.
\begin{equation}
D_i f = \partial_i f,
\end{equation}
while
\begin{equation}
D_i D_j f = \partial_i \partial_j f-\partial_k \Gamma_{ij}^k.
\end{equation}
In our special case, the only non-zero component of the symplectic connections are
\begin{equation}
\Gamma^{X}_{XY}= \frac{e}{3\hbar} \partial_X B(X), \hspace{0.3cm} 
\Gamma^{Y}_{XX}= -\frac{2e}{3\hbar} \partial_X B(X).
\end{equation}
We are now ready to introduce the Fedosov start-product, which is given by \cite{Dobrski,Blaschke}
\begin{align}
f \bigstar g = \hspace{3.3cm} \nonumber \\
 f g + \frac{i}{2} \theta^{ij} D_i f D_j g-\frac{1}{8} \theta^{kl} \theta^{ij} D_{(k} D_{i)} f D_{(l} D_{j)} g + \mathcal{O}(\theta^3),
\end{align}
where the brackets represent symmetrization. The higher order terms in $\theta$ explicitly contains the
symplectic Riemann tensor in Eq. (\ref{Riemann}), which is here omitted to simplify our discussion. Importantly,
when either $\theta^{ij}$ becomes constant or $R^i_{jkl}=0$, the above expression reduces to the standard Moyal star-product. Hence, we assume that for our choice of $B(X)$, some components of the corresponding symplectic curvature tensor do not vanish. In our case, this feature is satisfied by the following condition
\begin{equation}
R^X_{XXY}=\partial_X \Gamma^X_{XY}=\frac{e}{3\hbar} \partial^2_X B(X) \neq 0.
\end{equation}
Thus, the simplest configuration to get a curved symplectic manifold from a space-dependent magnetic field is simply given by
\begin{equation}\label{magnetic}
B(X)= a X^2+B_0,
\end{equation}
with $a$ and $B_0$ two real and positive constant parameters. Notice, this choice is also compatible with inversion symmetry and invertibility discussed previously.
Moreover, in analogy to the Moyal brackets, we can also construct the Fedosov brackets
\begin{align}
\{f, g\}_F=f \bigstar g - g \bigstar f.
\end{align}
Thus, the Fedosov's deformation quantization for symplectic manifolds with a compatible torsion-free symplectic connection (known also as Fedosov manifolds) represents the natural generalization of the standard deformation quantization that holds for symplectic vector spaces.
Moreover, it is straightforward to see that
\begin{align}
\{X, Y\}_F= i \theta^{XY}=\frac{i\,\hbar }{e(a X^2 + B_0)},
\end{align}
which agrees with the emergent noncommutative geometry of the quantum Hall states in presence of invertible and space-dependent magnetic fields.
We now explicitly calculate the Fedosov brackets for the plane waves in Eq.(\ref{planewaves}) with our choice of $B(X)$ in Eq.(\ref{magnetic})
\begin{align}
\{\rho(p_x,p_y), \rho(q_x,q_y)\}_F \approx \nonumber \\ \{\rho(p_x,p_y), \rho(q_x,q_y)\}_{M(\theta)} + \mathcal{O}(\theta^3),
\end{align}
where on the right we have denoted with the symbol $M(\theta)$ the Moyal product for which the square of the constant magnetic length $l_B$ is replaced by the space-dependent deforming parameter $\theta^{XY}$ (notice that in the above expression, the Moyal star product is approximated until $\mathcal{O}(\theta^2)$).
Because we are considering a weakly inhomogeneous magnetic field with $a \ll B_0$, we can expand Eq.(\ref{magnetic}) in the first order in $a$ obtaining
\begin{align}
\theta^{XY} \equiv \tilde{\theta}\approx l_B^2 - l_B^2 \left(\frac{a}{B_0}\right) X^{2} + \mathcal{O}(a^2).
\end{align}
Even in this simplified configuration, due to the space-dependent parameter $\tilde{\theta}$, the following quantity
\begin{align}
\langle \{\rho(p_x,p_y), \rho(q_x,q_y)\}_{M(\tilde{\theta})}\rangle\equiv \nonumber \\
\int dX dY\, \hat{\rho}(X,Y)\, \{\rho(p_x,p_y), \rho(q_x,q_y)\}_{M(\tilde{\theta})},
\end{align}
is not exactly equivalent to the commutator of two Fourier-transformed density operators like in the case of constant magnetic fields. In other words, for $B(X)$ in Eq.(\ref{magnetic}), the GMP algebra is not closed. This is a reminiscence of a similar problem that occurs in fractional Chern insulators that do not support nearly-flat topological bands with almost constant Berry curvature and quantum metric \cite{Roy,Sondhi2,Thomale2,Thomale,Ozawa2,Ozawa,Wang,Chen,Northe, Ledwith2,Estienne}.\\ 
This issue can be solved by taking the following configuration for the magnetic field
\begin{equation}\label{linear}
B(X)= a |X|+B_0,
\end{equation}
where the absolute value of the X coordinate guarantees the invariance under inversion symmetry. For the above choice of the magnetic field, all the components of the symplectic Riemann tensor $R^i_{jkl}$ are zero and this implies that 
we can choose vanishing symplectic connections and suitable Darboux local coordinates $X_D$ and $Y_D$ on the flat symplectic manifold such that the Fedosov start-product reduces to the Moyal star product \cite{Blaschke}. Thus, we get
\begin{equation}
\{X_D, Y_D\}_M= i l^2_B,
\end{equation}
 and in this second configuration the density operators satisfy the GMP algebra. 

 \section{Unimodular and background metrics}
  \noindent Through the Darboux coordinates for the case defined by Eq.(\ref{linear}), we can show now that a static Haldane's unimodular metric \cite{Haldane,Sondhi} naturally emerges. In fact, as shown in Ref.\cite{Dobrski}, in general, it is possible to define a compatible Riemmanian metric tensor $g_{ij}$ on any Fedosov manifold that satisfies the following weakly compatible condition
 \begin{equation}
 \Gamma^{k}_{ij}= \mathring{\Gamma}^{k}_{ij},
 \end{equation}
where 
\begin{equation}
\mathring{\Gamma}^{k}_{ij}=\frac{1}{2} g^{k l}(\partial_i g_{lj}+\partial_j g_{li}-\partial_l g_{ij}),
\end{equation}
which is the metric connection (Christoffel symbols), with $g^{k l}$ the inverse of the metric.
From the above condition, it is possible to derive the following constraint between and volume element associated to the metric and the symplectic volume related to the symplectic two-form $\omega_{ij}$
\begin{equation}\label{omega}
\sqrt{\det(g_{ij})}= C \sqrt{\det(\omega_{ij})},
\end{equation}
with C a positive constant. In our case, $g_{ij}=g_{ij}(X_D, Y_D)$ and $\omega_{ij}(X_D,Y_D)= \epsilon_{ij}/l_B^2$ such that the above expression becomes
\begin{equation}\label{metric}
\sqrt{\det(g_{ij})}=  c\, \frac{1}{l_B^2},
\end{equation}
which identifies an Euclidean unimodular metric (i.e. a metric that gives rise to a constant volume element) that  coincides with the static limit of the Haldane's unimodular metric. This metric should not be confused with the background metric, which is in general not unimodular and in this work, so far, it has been set equal to the flat metric of the plane.
Importantly the unimodularity is clearly lost for the case related to Eq.(\ref{magnetic}), in which the magnetic field is quadratic in the X coordinate. 
Notice that Eq.(\ref{metric}) is similar (but not equivalent) to the relation between the momentum-space quantum metric $G_{ij}(p_x,p_y)$ and the the Berry curvature $\Omega_{xy}(p_x,p_y)$ in ideal Chern insulators \cite{Roy,Sondhi2,Thomale2,Thomale,Ozawa2,Ozawa,Wang,Chen,Northe,Ledwith2,Estienne}
\begin{equation}
\sqrt{\det(G_{ij})}=\frac{1}{2}  |\Omega_{xy}|.
\end{equation}
For a single Landau level, we have $\Omega_{xy}(p_x,p_y)=l_B^2$ \cite{Ozawa}. This clearly shows a duality between the symplectic compatible metric in real space and the quantum metric in momentum space.\\
Finally, we briefly discuss the case in with we switch on a curved background $\hat{g}_{ij}$ by keeping a generic non-uniform magnetic field. It has been shown that the LLs are still dispersionless and degenerate iff \cite{Alicki,Klevtsov3, Klevtsov,Wiegmann}
\begin{equation}\label{metric2}
\sqrt{\det(\hat{g}_{ij})}\propto B(X,Y).
\end{equation}
By combining the above relation with the general results in Eqs.(\ref{omega2}) and (\ref{omega}), we then obtain
\begin{equation}\label{metric3}
\sqrt{\det(\hat{g}_{ij})} \propto \sqrt{\det(g_{ij})},
\end{equation}
which represents a proportionality relation (up to a constant factor) between the background and unimodular metrics.
A special but straightforward solution is given by
\begin{equation}\label{metric4}
\hat{g}_{ij} =\Omega\, g_{ij},
\end{equation}
with $\Omega$ a real constant factor.
In this way, we have shown that through symplectic geometry the LLs can be completely characterized by two distinct metrics. This result is in agreement with the general idea to formalize quantum Hall fluids in terms of bimetric geometry \cite{Bradlyn,Gromov}. 
However, Eq.(\ref{metric3}) tells us that for dispersionless and degenerate LLs in curved geometry, the unimodularity of the Haldane's metric is lost unless we artificially consider some special two-dimensional spaces characterized by unimodular background metric tensors.
 
\section{Conclusions and outlook}
\noindent Summarizing, in this work, we have provided some physical conditions to define noncommutative geometry in the quantum Hall effect in presence of inhomogeneous magnetic fields. We have discussed two different configurations for the magnetic fields. Some issues arise to define a closed GMP algebra for the momentum-space density operators in the case the symplectic curvature tensor is not zero. This poses some new challenge to define fractional quantum Hall states in general non-uniform magnetic fields. Nevertheless, this problem can be solved when the magnetic field is linear in the absolute value of the space coordinates as showed at the end of the previous section. Moreover, we have shown that there exists a natural Riemmanian metric that is compatible with the symplectic two-form and can be related to the Haldane's unimodular metric. Thus, we believe that deformation quantization and symplectic geometry provide a powerful approach to understand quantum Hall fluids even beyond constant magnetic fields. About possible future directions, it would be interesting to employ Fedosov's deformation quantization and symplectic geometry in bilayer Hall systems under inhomogeneous magnetic fields \cite{Hasebe3}, in the supersymmetric generalization of fractional quantum Hall states \cite{Hasebe2} and in suitable three-dimensional topological fluids where the presence of Nambu brackets (which generalize Poisson brackets) give rise to a non-associative geometry \cite{Neupert}.
	\vspace{0.5cm}

\noindent {\bf Acknowledgments: }
The author is pleased to acknowledge discussions with Blagoje Oblak and Dimitra Karabali.

	
		\bibliography{references}

\end{document}